\documentclass[conference]{IEEEtran}

\usepackage{cite}

\usepackage{graphicx}
\usepackage{subfigure}
\usepackage{boxedminipage}

\ifCLASSINFOpdf
\else
\fi
\usepackage[cmex10]{amsmath}
\usepackage{amsmath}
\usepackage{algorithmic}

\usepackage{times,color}
\usepackage[final,ulem=normalem]{changes}
\definechangesauthor{RW}{blue}
\definechangesauthor{AW}{red}
\definechangesauthor{TC}{orange}

\begin{document}
%
\title{A Rate-Compatible Sphere-Packing Analysis of Feedback Coding with Limited Retransmissions}

\author{\IEEEauthorblockN{Adam R. Williamson, Tsung-Yi Chen and Richard D. Wesel}
\IEEEauthorblockA{Department of Electrical Engineering\\
University of California, Los Angeles\\
Los Angeles, California 90095\\
Email: adamroyce@ucla.edu; tychen@ee.ucla.edu; wesel@ee.ucla.edu}
}

\maketitle

\begin{abstract}
Recent work by Polyanskiy et al. and Chen et al. has excited new interest in using feedback to approach capacity with low latency.   Polyanskiy showed that feedback identifying the first symbol at which decoding is successful allows capacity to be approached with surprisingly low latency. This paper uses Chen's rate-compatible sphere-packing (RCSP) analysis to study what happens when symbols must be transmitted in packets, as with a traditional hybrid ARQ system, and limited to relatively few (six or fewer) incremental transmissions.
  
Numerical optimizations find the series of progressively growing cumulative block lengths that enable RCSP to approach capacity with the minimum possible latency.  RCSP analysis shows that five incremental transmissions are sufficient to achieve 92\% of capacity with an average block length of fewer than 101 symbols on the AWGN channel with SNR of 2.0 dB.
  
The RCSP analysis provides a decoding error trajectory that specifies the decoding error rate for each cumulative block length. Though RCSP is an idealization, an example tail-biting convolutional  code matches the RCSP decoding error trajectory and achieves 91\% of capacity with an average block length of 102 symbols on the AWGN channel with SNR of 2.0 dB.  We also show how RCSP analysis can be used in cases where packets have deadlines associated with them (leading to an outage probability). 
%
%
\end{abstract}


%
\IEEEpeerreviewmaketitle

%
%
\section{Introduction}

Though Shannon showed in 1956 \cite{Shannon_IT_1956} that noiseless feedback does not increase the capacity of memoryless channels, feedback's other benefits have made it a staple in modern communication systems. Feedback can simplify the encoding and decoding operations and has been incorporated into incremental redundancy (IR) schemes proposed as early as 1974 \cite{Mandelbaum_IT_1974}. Hagenauer's work on rate-compatible punctured convolutional (RCPC) codes allows the same encoder to be used in various channel conditions and uses feedback to determine when to send additional coded bits \cite{Hagenauer_TCOM_1988}.  The combination of IR and hybrid ARQ (HARQ) continues to receive attention in the literature \cite{Denic_ISIT_2011, Soljanin_ITW_2009, Liu_Soljanin_ITW_2003} and industry standards such as 3GPP. 



Although it cannot increase capacity in point-to-point channels, the information-theoretic benefit of feedback for reducing latency through a significant improvement in the error exponent has been well understood for some time. (See, for example, \cite{Dobrushin_1962,Burnashev_1976,Burnashev_1980,Schalkwijk_1966_1}.)  Recent work \cite{Polyanskiy_IT_2011_NonAsym,Chen_2011_ICC, Chen_2011_ITA} casts the latency benefit of feedback in terms of block length rather than error exponent, generating new interest in the practical value of feedback for approaching capacity with a short average block length.

Polyanskiy et al. provided bounds for the maximum rate that can be accomplished with feedback for a finite block length \cite{Polyanskiy_IT_2011_NonAsym} and also demonstrated the energy-efficiency gains made possible by feedback \cite{Polyankskiy_IT_2011_MinEnergy}.  In the variable-length feedback with termination (VLFT) scheme, \cite{Polyanskiy_IT_2011_NonAsym} uses an elegant, single, ``stop feedback'' symbol (that can occur after any transmitted symbol) that facilitates the application of Martingale theory to capture the essence of how feedback can allow a variable-length code to approach capacity.  A compelling example from \cite{Polyanskiy_IT_2011_NonAsym} shows that for a binary symmetric channel with capacity 1/2, the average block length required to achieve 90\% of the capacity is smaller than 200 symbols.  

For practical systems such as hybrid ARQ, the ``stop feedback'' symbol may only be feasible at certain symbol times because these systems group symbols together for transmission in packets, so that the entire packet is either transmitted or not.   In \cite{Chen_2011_ICC, Chen_2011_ITA},  Chen et al. used a code-independent rate-compatible sphere-packing (RCSP) analysis to quantify the latency benefits of feedback in the context of such grouped transmissions.  Chen et al. focused on the AWGN channel and also showed that capacity can be approached with surprisingly small block lengths, similar to the results of \cite{Polyanskiy_IT_2011_NonAsym}. 

Using the RCSP approach of Chen et al. as its foundation, this paper introduces an optimization technique and uses it to explore how closely one may approach capacity with only a handful of incremental transmissions.  For a fixed number of information bits $k$ and a fixed number of maximum transmissions $m$ before giving up to try again from scratch, a numerical optimization determines the block lengths of each incremental transmission to maximize the expected throughput.  We consider only $m$$\le$$6$ and show that this is sufficient to achieve more than 90\% of capacity while requiring surprisingly small block lengths similar to those achieved by Polyanskiy et al. and Chen et al.

While RCSP is an idealized scheme, it provides meaningful guidance for the selection of block lengths and the sequence of target decoding error rates, which we call the decoding error trajectory.  A 1024-state rate-compatible punctured tail-biting convolutional code using the block lengths determined by our RCSP optimization technique achieves the RCSP decoding error trajectory and essentially matches the throughput and latency performance of RCSP for $m$$=$$5$ transmissions. Our results, like those of Polyanskiy et al. and Chen et al., assume that the receiver is able to recognize when it has successfully decoded.  The additional overhead of, for example, a cyclic redundancy check (CRC) has not been included in the analysis.  Longer block lengths would be required to overcome this overhead penalty.

The paper is organized as follows: Section~\ref{sec:analysis} reviews the RCSP analysis. Section \ref{sec:algo} describes the RCSP numerical optimization used to determine transmission lengths and shows the throughput vs. latency performance  achieved by using these transmission lengths for up to six rate-compatible transmissions.  This performance is compared with a version of VLFT scheme proposed by Polyanskiy et al. Section \ref{sec:sims} introduces the decoding error trajectory and shows how RCSP performance can be matched by a real convolutional code using the transmission lengths identified in the previous section.  Section \ref{sec:outage} shows how the RCSP analysis can be applied to scenarios that involve strict latency and outage probability constraints.  Section \ref{sec:conc} concludes the paper.

%
%
\section{Rate-Compatible Sphere-Packing (RCSP)}
\label{sec:analysis}

\subsection{Review of Sphere-Packing}

To review the sphere-packing analysis presented in \cite{Chen_2011_ICC,Cover_Thomas_1991} for a memoryless AWGN channel, consider a codebook of size $2^{k}$  that maps $k = NR_c$ information symbols into a length-$N$ codeword with rate $R_c$. The channel input and output can be written as:
	\begin{equation}
	Y = X(j) + Z, j \in 1,2, \dots ,2^k,
	\end{equation}
where $Y$ is the output (received word), $X(j)$ is the codeword of the $j$th message, and Z is an $N$-dimensional i.i.d. Gaussian vector.
Let the received SNR be $\eta$ and assume without loss of generality that each noise sample has unit variance. The average power of received word $Y$ is then $N(1+\eta)$. As in \cite{Chen_2011_ICC}, the largest possible squared decoding radius $r^2$ assuming that the decoding spheres occupy all available volume is
	\begin{equation}
	r^2 = N(1 + \eta)~2^{-2k/N}.
	\end{equation}
A bounded-distance decoder declares any message within a distance $r$ of codeword $X(j)$ to be message $j$. Otherwise, a decoding error is declared. Because the sum of the squares of the $N$ Gaussian noise samples obeys a chi-square distribution with $N$ degrees of freedom, the probability $P(\zeta)$ of decoding error associated with decoding radius $r$ is 
	\begin{equation}
	P(\zeta) = P\bigg(\sum\limits_{\ell = 1}^{N} z_\ell^2 > r^2 \bigg) = 1- F_{\chi_{N}^2}(r^2),
	\label{eqn:Pzeta}
	\end{equation}
where the $z_\ell$ are standard normal distributed random variables with zero mean and unit variance and $F_{\chi_{N}^2}(x)$ is the CDF of a chi-square distribution with $N$ degrees of freedom.

\subsection{Sphere-Packing for Rate-Compatible Transmissions}

The idea of RCSP is to assume that sphere-packing performance can be achieved by each transmission in a sequence of rate-compatible transmissions.  Thus  the idealized sphere-packing analysis is applied to a modified incremental redundancy with feedback (MIRF) scheme as described in \cite{Chen_2011_ICC}. MIRF works as follows: $k$ information symbols are coded with an initial block length $N_1=I_1$. If the receiver cannot successfully decode, the transmitter will receive a NACK and send $I_2$ extra symbols. The decoder attempts to decode again using all received symbols for the current codeword, i.e., with block length $N_2 = I_1 + I_2$. The process continues for $i=3,\dots,m$. The decoded block length $N_j$ is $I_1 +I_2 + \dots + I_j$ and the code rate is $R_j = k/N_j$.  If decoding is not successful after $m$ transmissions, the decoder discards the $m$ transmissions and the process begins again with the transmitter resending the $I_1$ initial symbols. This scheme with $m$=$1$ is standard ARQ.

The squared decoding radius of the $j$th cumulative transmission is 
	\begin{equation}
	r_j^2 = N_j(1 + \eta) ~ 2^{-2k/N_j},
	\label{eqn:r_i}
	\end{equation}
and the marginal probability of decoding error $P(\zeta_j)$ associated with decoding radius $r_j$ is
	\begin{equation}
	P(\zeta_j) = P\bigg(\sum\limits_{\ell = 1}^{N_j} z_\ell^2 > r_j^2 \bigg),
	\end{equation}
where the $z_{\ell}$ are standard normal distributed random variables with zero mean and unit variance. 

However, this marginal probability is not what is needed.  The probability of a decoding error in the $j$th transmission depends on previous error events.  Indeed, conditioning on previous decoding errors $\zeta_1, \ldots, \zeta_{j-1}$ makes the error event  $\zeta_{j}$ more likely than the marginal distribution would suggest.  The joint probability $P(\zeta_1,\dots,\zeta_j)$ is
	\begin{align}
	P(&\zeta_1,\dots,\zeta_j)=P\bigg(\bigcap \limits _{i=1}^{j} \zeta_i\bigg) \nonumber\\
	=& \int_{r_1^2}^{\infty} \int_{r_2^2-t_1}^{\infty} \dots \int_{r_{j-1}^2- \sum_{i=1}^{j-2}t_i}^{\infty}  f_{\chi_{I_1}^2}(t_1) \dots f_{\chi_{I_{j-1}}^2}(t_{j-1}) \times\nonumber\\
&\left(1-F_{\chi_{I_{j}}^2} \bigg (r_j^2 - \sum_{i=1}^{j-1}t_i\bigg ) \right) dt_{j-1} \dots dt_1.
	\label{eqn:Pzeta_i}
	\end{align}

We compute the expected number of channel uses (i.e., latency or average block length) $\lambda$ by summing the incremental transmission lengths $I_i$ weighted by the probability of error in the prior cumulative transmission and dividing by the probability of success by the last ($m$th) transmission (as in ARQ), according to
	\begin{equation}
	\lambda = \frac{I_1+\sum\limits_{i=2}^{m} I_i P\bigg(\bigcap \limits _{j=1}^{i-1} \zeta_j\bigg)}{1 - P\bigg(\bigcap \limits_{j=1}^{m} \zeta_j\bigg)}	\label{eqn:lambda_tau}.
	\end{equation}
This expression does not consider delay due to decoding operations. The corresponding throughput $R_t$ is given by
	\begin{equation}
	R_t = \frac{k}{\lambda} = \frac{k \left({1 - P\bigg(\bigcap \limits_{j=1}^{m} \zeta_j\bigg)}\right)}{I_1+\sum\limits_{i=2}^{m} I_i P\bigg(\bigcap \limits _{j=1}^{i -1}\zeta_j\bigg)}.
	\label{eqn:Rt}
	\end{equation}

\section{Choosing $I_i$ Values to Maximize Throughput}
\label{sec:algo}
\subsection{Selecting $I_1$ for the $m=1$ (ARQ) Special Case}
\label{sec:optARQ}
In the special case of  $m$=$1$ (when only the initial transmission of length $I_1$ is ever transmitted), MIRF is ARQ.  In this case the expected number of channel uses given by (\ref{eqn:lambda_tau}) can be simplified as follows (with $r_1^2= \frac{k/R_c (1 + \eta)}{2^{2R_c}}$):
	\begin{equation}
	\lambda_{ARQ} = \frac{I_1}{1 - P\left(\zeta_1\right)} = \frac{I_1}{F_{\chi_{I_1}^2}(r_1^2)},\\
	\label{eqn:lambda_ARQ}
	\end{equation}
which yields an expected throughput of 
	\begin{equation}
	{R_t}_{ARQ} = {(k/I_1) F_{\chi_{I_1}^2}(r_1^2)} = R_c F_{\chi_{k/R_c}^2}(r_1^2).\\
	\label{eqn:Rt_ARQ}
	\end{equation}

If we fix the number of information bits $k$, (\ref{eqn:Rt_ARQ}) becomes a quasiconcave function of the initial code rate $R_c=k/I_1$, allowing the optimal code rate $R_c^{opt}$, which maximizes the throughput $R_t$ for a given $k$, to be found numerically \cite{Boyd_2004_CO}.  Fig. \ref{fig:latVthroughput_A} plots the maximum achievable throughput in the $m$$=$$1$ (ARQ) RCSP scheme as the red (diamond markers) curve. 

\subsection{Optimizing $I_i$ Values for $m>1$}
\label{sec:subalgo}

In \cite{Chen_2011_ICC}, Chen et al. demonstrated one specific RCSP scheme with ten transmissions that could approach capacity with low latency.  Specifically, the transmission lengths were fixed to $I_1$$=$$64$ and $I_2, \ldots, I_{10}$$=$$10$, while $k$ was varied to maximize throughput. This paper builds on the intuition of the ARQ case presented above.   Both $k$ and the number of transmissions $m$ are fixed, and a search identifies the set of transmission lengths $I_{i}$ that maximizes throughput.   We seek to identify approximately how  much throughput can be achieved using feedback with a small number of incremental transmissions, specifically $m$$\le$$6$.   Furthermore, we seek insight into what the transmission lengths should be and what decoding error rates allow the sequence of transmissions to be most efficient.

For $m$$>$$1$, identifying the transmission lengths $I_{i}$ which minimize the latency $\lambda$ in (\ref{eqn:lambda_tau}) is not straightforward due to the joint decoding error probabilities in (\ref{eqn:Pzeta_i}).  However, the restriction to a small $m$ allows exact computation of  \eqref{eqn:Pzeta_i} in Mathematica, avoiding the approximations of\cite{Chen_2011_ICC}.  To reflect practical constraints, we restrict the lengths $I_i$ to be integers. 


\begin{figure}
  \centering
    \subfigure[RCSP Analysis]{ 
   \label{fig:latVthroughput_A}
    \scalebox{0.5}{\includegraphics{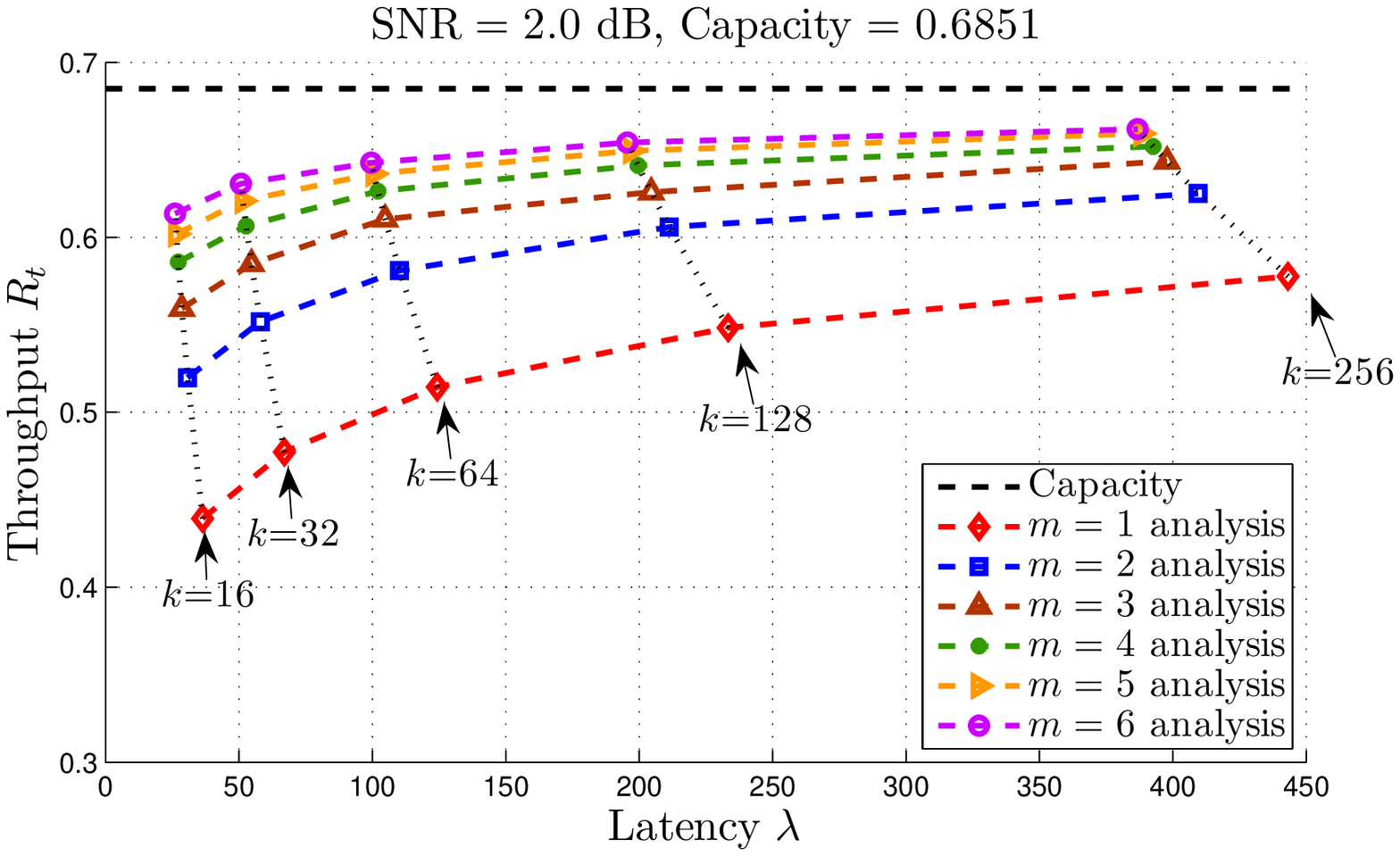}}
    }
    \hspace{0.0in}
    \subfigure[$m=5$ RCSP Analysis, Simulations, and VLFT Comparison]{
   \label {fig:latVthroughput_B}
    \scalebox{0.5}{\includegraphics{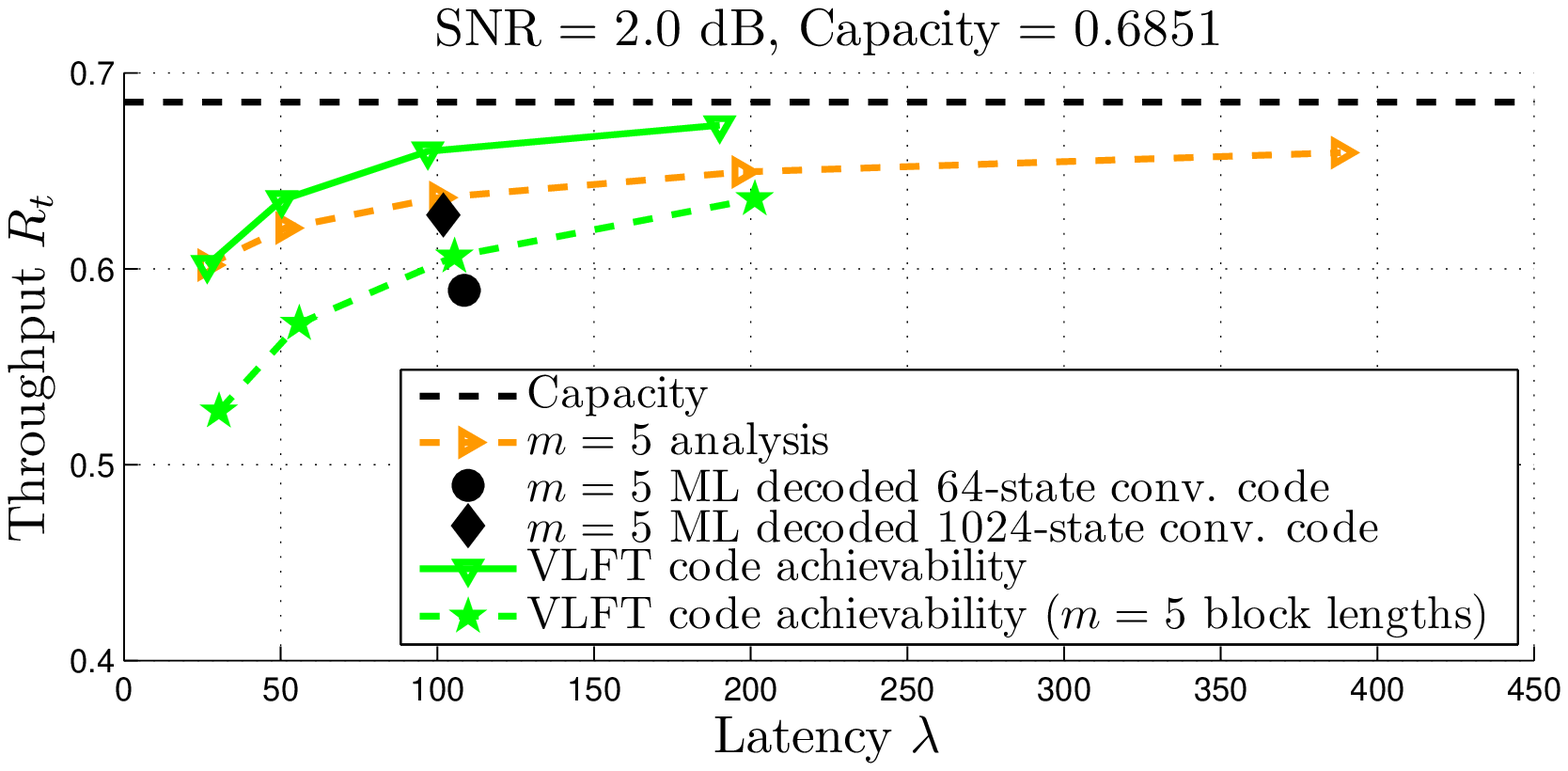}}
    }
    \hspace{0.0in}
    \vspace{-6pt}
  \caption{Throughput vs. latency for RCSP  with $m$ rate-compatible transmissions $m\in \{1, \dots, 6\}$ with transmission lengths $I_i$ identified by RCSP optimization.  Also shown are convolutional code simulations, VLFT and a constrained version of VLFT that uses the RCSP $m$=$5$ block lengths.}
  \vspace{-16pt}
\end{figure}

The computational complexity of \eqref{eqn:Pzeta_i}, which increases with the transmission index $j$, forces us to limit attention to a well-chosen subset of possible transmission lengths. Thus, our present results may be considered as lower bounds to what is possible with a fully exhaustive optimization. Fig. \ref{fig:latVthroughput_A} shows the throughput vs. latency performance achieved by RCSP for $m \in \{1, \ldots 6\}$ on an AWGN channel with SNR 2.0 dB. As $m$ is increased, each additional retransmission brings the expected throughput $R_t$ closer to the channel capacity, though with diminishing returns. The points on each curve in Fig. \ref{fig:latVthroughput_A} represent values of $k$ ranging from 16 to 256 information bits. Fig. \ref{fig:latVthroughput_A} shows, for example, that by allowing up to four retransmissions ($m$$=$$5$) with $k$$=$$64$, RCSP can achieve 91\% of capacity with an average block length of 102 symbols. Similar results are obtained for other SNRs. 

VLFT achievability results for the AWGN channel based on \cite{Polyanskiy_IT_2011_NonAsym} are shown in Fig. \ref{fig:latVthroughput_B} for comparison.  Both the original VLFT scheme, in which the transmission may be ended after any symbol, and a constrained version of VLFT using the same block lengths and feedback structure as $m$=$5$ RCSP are presented.  The original VLFT closely approaches capacity with a latency on the order of 200 symbols.  RCSP is unable to match VLFT because the overall RCSP transmission can only be terminated after one of the $m$ incremental transmissions completes.  If VLFT is constrained in the same way, its performance is initially worse than RCSP because random coding does not achieve ideal sphere packing with short block lengths.  At an average latency of 200, constrained VLFT  performance becomes similar to the comparable RCSP scheme.  

The VLFT achievability curve evaluates \cite[Theorem 10]{Polyanskiy_IT_2011_NonAsym} using the upper bound of (162) with i.i.d. Gaussian inputs with average power equal to the power constraint $\eta$.  Such codebooks will sometimes violate the 2 dB power constraint.  To address this, the average power should be slightly reduced and codebooks violating the power constraint should be purged, which will lead to a small performance degradation. Alternatively, codebooks or even codewords can be constrained to meet the power constraint with equality. Further analysis of VLFT codes more carefully considering the power constraint for the AWGN channel will be the subject of future work.

%
%
\section{Comparison of RCSP and Convolutional Codes}
\label{sec:sims}

RCSP makes the rather optimistic assumption that a family of rate-compatible codes can be found that performs, at each rate, equally well as codes that pack decoding spheres so well that they use all of the available volume. A variety of well-known upper bounds on the packing density $\phi$ indicate that the maximum packing density decreases as the dimension $n$ increases (e.g., $\phi~\leq~(n/e)~2^{-n/2}$) \cite{Torquato_2006_NCL}, making such codes difficult to find. However, we show in this section that a rate-compatible tail-biting convolutional code can indeed match the performance of RCSP, at least for $m$$=$$5$.  

\subsection{Two Convolutional Codes}
We consider two rate $1$/$3$ convolutional codes from \cite{Lin_2004_ECC}: a 64-state code with generator polynomial ($g_1$,$g_2$,$g_3$)=($133,171,165$) and a 1024-state code with ($g_1$,$g_2$,$g_3$)=($3645,2133,3347$), where the generator notation is octal. High rate codewords are created by pseudorandom rate-compatible puncturing of the rate $1$/$3$ mother codes. We restrict our attention to tail-biting implementations of these convolutional codes because the throughput efficiency advantage is important for the relatively small block lengths we consider. Simulations compare the performance of these two codes in the MIRF setting for the AWGN channel with SNR 2 dB, as shown in Fig. \ref{fig:latVthroughput_B}.  The simulations presented here focus on the $k$$=$$64$ case.
\begin{table}
\begin{center}
  \caption{Optimal RCSP transmission lengths for $m=5$ and SNR 2 dB.} 
\begin{tabular}{| c | c | c | c | c | c |}
\hline
  $k$ & $I_1$ & $I_3$ & $I_3$ & $I_4$ & $I_5$ \\
  \hline
  16 &   19 & 4 &    4     & 4     & 8 \\
  \hline
  32 &   38 &    8 &    8  &   8 &   12 \\
  \hline
    64   & 85 &   12   &  8  &  12   & 16 \\
  \hline
   128  & 176  &  14  &  14   & 14  &  28 \\
  \hline
   256 &  352  &    24 &   24  &  24  &  48 \\
  \hline
\end{tabular}
\label{table:m5_steps}
\end{center}
  \vspace{-10pt}
\end{table}

\begin{figure}
  \centering
    	\scalebox{0.47}{\includegraphics{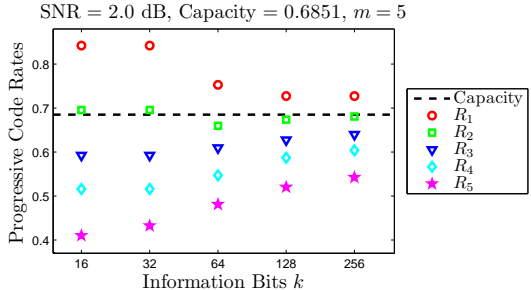}}
    \hspace{0.1in}
    \vspace{-4pt}
\caption {Rates of the five cumulative transmissions for $m=5$ and an AWGN channel with SNR 2 dB for $k$=$16$, $32$, $64$, $128$, and $256$.}
\label{fig:rates_vs_k}
  \vspace{-12pt}
\end{figure}

The transmission lengths $I_i$ used in the simulations are those identified by the RCSP optimization. Table \ref{table:m5_steps} shows the results of the $m$$=$$5$ optimization (i.e., the set of lengths $I_i$ found to achieve the highest throughput).  Thus our simulations used $I_1$$=$$85$, $I_2$$=$$12$, $I_3$$=$$8$, $I_4$$=$$12$, $I_5$$=$$16$.  The induced code rates of the cumulative blocks are $64$/$85$=$0.753$, $64$/$97$=$0.660$, $64$/$105$=$0.610$, $64$/$117$=$0.547$ and $64$/$133$=$0.481$.  Fig. \ref{fig:rates_vs_k} shows these rates as well as the rates for other values of $k$ according to the RCSP optimization for $m=5$.  Note that for every value of $k$ the initial code rate is above the channel capacity of 0.6851. This is the benefit of feedback: it allows the decoder to capitalize on favorable noise realizations by attempting to decode early, instead of needlessly sending additional symbols.

\subsection {Decoding Error Trajectory Comparison}

\begin{figure}
  \centering
    	\scalebox{0.47}{\includegraphics{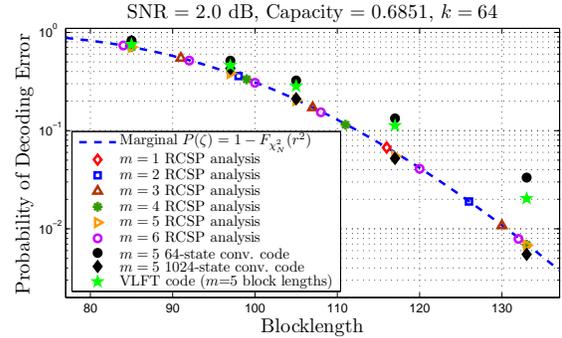}}
    \hspace{0.1in}
    \vspace{-4pt}
\caption {A comparison of the decoding error trajectories of RCSP, simulated ML-decoded convolutional codes and VLFT, for $k$=$64$.}
\label{fig:PrErrorM5K64}
  \vspace{-12pt}
\end{figure}

The RCSP optimization also computes the joint decoding error probabilities of \eqref{eqn:Pzeta_i}, which we call the ``decoding error trajectory''.  If we can find a rate-compatible family that achieves this decoding error trajectory, then we can match the RCSP performance.  Fig. \ref{fig:PrErrorM5K64} shows the $k$=$64$ decoding error trajectories for  the RCSP cases studied  in Figs. 1 and 2 (shown as $m$ discrete points in Fig. \ref{fig:PrErrorM5K64} for each value of $m \in \{1, \ldots, 6\}$) and for constrained VLFT for $m=5$ and 64-state and 1024-state convolutional code simulations for $m$=$5$.  The dashed line represents the marginal probability of error for a sphere-packing codebook as in \eqref{eqn:Pzeta} which was recognized in \cite{ChenITW2012} as a tight upper bound for the joint probabilities of error given by \eqref{eqn:Pzeta_i}.   This tight upper bound can serve as a performance goal for practical rate-compatible code design across a wide range of block lengths.

While the 64-state code is not powerful enough to match RCSP performance, the 1024-state code closely follows the RCSP trajectory for $m=5$.  Thus there exist practical codes, at least in some cases, that achieve the idealized performance of RCSP.  Indeed,  Fig. \ref{fig:latVthroughput_B} plots the $(\lambda,R_t)$ points of the two convolutional codes, demonstrating that the 1024-state code achieves 91\% of capacity with an average latency of 102 symbols, almost exactly coinciding with the RCSP point for $m$$=$$5$ and $k$$=$$64$. The convolutional code's ability to match a mythical sphere-packing code is due to maximum likelihood (ML) decoding, which has decoding regions that completely fill the multidimensional space (even in high dimensions).

\subsection{Caveat}
These simulation results  assume that the receiver is able to recognize when it has successfully decoded. This same assumption is made by the RCSP analysis, the VLFT scheme of Polyanskiy et al., and the MIRF scheme of Chen et al.  While this assumption does not undermine the essence of this demonstration of the power of feedback, its practical and theoretical implications must be reviewed carefully, especially when very short block lengths are considered. An important practical  implication is that the additional overhead of a CRC required to avoid undetected errors will drive real systems to somewhat longer block lengths than those presented here. This will affect the choice of error control code. An important implication is that this analysis cannot be trusted if the block lengths become too small.   This assumption allows block errors to become block erasures at no cost.  Consider the binary symmetric channel (BSC): If the block length is allowed to shrink to a single bit, then this assumption turns  the zero capacity BSC with transition probability $1/2$ into a binary erasure channel with probability $1/2$, which has a capacity of $1/2$ instead of zero. Both the practical and theoretical problems of this assumption diminish as block length grows.  However, a quantitative understanding of the cost of knowing when decoding is successful and how that cost changes with block length is an important area for future work.

\section{RCSP with Latency and Outage Constraints}
\label{sec:outage} 

MIRF has an outage probability of zero because it never stops trying until a message is decoded correctly.  With slight modifications, the MIRF scheme and the RCSP transmission length optimization can incorporate strict constraints on latency (so that the transmitter gives up after $m$ transmissions) and outage probability (which would then be nonzero).

To handle these two new constraints, we restrict $P(\zeta_1,\dots,\zeta_m)$ to be less than a specified $p_{\text{outage}}$. Without modifying the computations of $P(\zeta_1,\dots,\zeta_j)$ in (\ref{eqn:Pzeta_i}), the optimization is adapted to pick the set of lengths that yields the maximum throughput s.t. $P(\zeta_1,\dots,\zeta_m) \leq p_{\text{outage}}$.  When there is a decoding error after the $m$th transmission, the transmitter declares an outage event and proceeds to encode the next $k$ information bits. This scheme is suitable for delay-sensitive communications, in which data packets are not useful to the receiver after a deadline has passed.  The expected number of channel uses $\lambda$ is now given by
	\begin{equation}
	\lambda = I_1+\sum\limits_{i=2}^{m} I_i P\bigg(\bigcap \limits _{j=1}^{i-1} \zeta_j\bigg).
		\label{eqn:lambda_tau_noRep}
	\end{equation}
The expected throughput $R_t$ is again given by (\ref{eqn:Rt}). 
%
%
\section{Conclusion}
\label{sec:conc}

The purpose of this paper is to bring the information theory of feedback and the communication practice of feedback closer together.  Beginning with the idealized notion of rate-compatible codes with decoding spheres that completely fill the available volume, the paper eventually demonstrates a convolutional code with performance strikingly similar to the ideal rate-compatible sphere-packing (RCSP) codes.

An optimization based on RCSP identifies the highest throughput possible for a fixed $k$ and $m$.  This optimization provides the lengths of the initial and subsequent transmissions and the sequence of decoding error probabilities or ``decoding error trajectory'' that characterize the throughput-maximizing performance. The RCSP decoding error trajectories computed in this paper are tightly bounded by the marginal error probability of sphere packing.  Designing a code with a similar error trajectory will thus yield comparable latency performance.  RCSP predictions and simulation results agree in demonstrating that feedback permits 90\% of capacity to be achieved with about 100 transmitted symbols assuming that the decoder knows when it has decoded correctly.  However, the implications of this assumption for short block lengths warrant further investigation.

VLFT performance shows that if the transmission could be stopped at any symbol (rather than only at the end of each incremental transmission) capacity is closely approached with an average latency of 200 symbols, but a more careful analysis of VLFT in light of the AWGN power constraint is warranted.

\section*{Acknowledgement}

The authors would like to thank Yury Polyanskiy for helpful conversations regarding the VLFT analysis.



%

\bibliographystyle{IEEEtran}
{\bibliography{ISIT2012_AW_bib}}

\newpage

\end{document}